\documentclass{article}
\usepackage{amssymb}
\usepackage{textcomp}
\usepackage{graphicx}

\newcommand{\bi}{\begin{itemize}}
\newcommand{\ei}{\end{itemize}}
\newcommand{\beq}{\begin{equation}}
\newcommand{\eeq}{\end{equation}}
\newcommand{\ben}{\begin{enumerate}}
\newcommand{\een}{\end{enumerate}}
\newcommand{\bdm}{\begin{displaymath}}
\newcommand{\edm}{\end{displaymath}}
\newcommand{\bea}{\begin{eqnarray}}
\newcommand{\eea}{\end{eqnarray}}
\newcommand{\etal}{{\it et al.}}
\newcommand{\ie}{{\it i.e.}}
\newcommand{\eg}{{\it e.g.}}

\newcommand{\lag}{{\mathcal L}}

\newcommand{\bino}{{\widetilde B}}
\newcommand{\wino}{{\widetilde W}}
\newcommand{\higgsino}{{\widetilde H}}

\newcommand{\re}{{\rm Re}}

\newcommand{\superlfv}{{\tt SuperLFV}\,\,}

\begin{document}

\begin{titlepage}

\thispagestyle{empty}
\setcounter{page}{0}
\def\thefootnote{\fnsymbol{footnote}}

\begin{center}
{\Large\bf SuperLFV:\\
An SLHA tool for lepton flavor violating observables in supersymmetric models}
\end{center}

\large

\vspace{.15in}
\begin{center}

Brandon Murakami\footnote{\tt bmurakami@ric.edu}

\small {\it Physical Sciences Department, Rhode Island College, Providence, RI 02908}

\end{center}
 
\vspace{0.15in}
 
\begin{abstract}

We introduce {\tt SuperLFV}, a numerical tool for calculating low-energy observables that exhibit charged lepton flavor violation (LFV) in the context of the minimal supersymmetric standard model (MSSM).  As the Large Hadron Collider and MEG, a dedicated $\mu^+\to e^+\gamma$ experiment, are presently acquiring data, there is need for tools that provide rapid discrimination of models that exhibit LFV.  \superlfv accepts a spectrum file compliant with the SUSY Les Houches Accord (SLHA), containing the MSSM couplings and masses with complex phases at the supersymmetry breaking scale.  In this manner, \superlfv is compatible with but divorced from existing SLHA spectrum calculators that provides the low energy spectrum.  Hence, input spectra are not confined to the LFV sources provided by established SLHA spectrum calculators.  Input spectra may be generated by personal code or by hand, allowing for arbitrary models not supported by existing spectrum calculators.  

\end{abstract}

\end{titlepage}

\tableofcontents

\section{Introduction}

The era of the Large Hadron Collider is expected to be a data driven era.  This statement is strengthened by a mix of numerous non-collider experiments, such as cold dark matter searches and flavor-centric probes, and near-future $B$ factories.  As experimental data becomes available in the form of measurements and null results, the value of the ability for rapid model discrimination increases.  The SUSY Les Houches Accord (SLHA) is a protocol for interfacing the input and output of numerical tools that perform calculations within supersymmetric models \cite{Skands:2003cj}.  From here, all usage of the acronym ``SLHA'' will refer to the updated SLHA2 format and beyond.  \superlfv is a contribution to the growing library of SLHA tools, with an initial primary focus on calculating low-energy observables with charged lepton flavor-violation (LFV) \cite{Marciano:2008zz}.

The standard model (SM) with massless neutrinos prohibits LFV, due to an accidental symmetry -- accidental in the sense that no known reason protects the symmetry.  Generically, extensions of the SM will break this symmetry and exhibit new sources of flavor-changing neutral currents (FCNC).  While it is observed that quarks exhibit FCNC and neutrinos oscillate flavor, thus far, charged leptons have not revealed flavor violation.  The fact that neutrinos oscillate establishes {\it charged} LFV as a theoretical prediction.  However, this prediction, given reasonable assumptions of unmeasured neutrino parameters, yields an undetectable $\mu\to e\gamma$ branching ratio of several orders beyond ~$10^{-50}$.  Hence any detection of $\mu\to e\gamma$ is an unmistakeable signature of new physics.  For this reason, despite whether the branching ratio BR($\mu\to e\gamma$) is measured or a stronger upper limit is placed, BR($\mu\to e\gamma$) should be considered a standard benchmark for evaluating models.

Supersymmetric models remain a primary focus of the field and, like any extension of the SM, generically exhibits LFV.  The first calculation of the $
\mu\to e\gamma$ branching ratio in a supersymmetric model was demonstrated in ref.~\cite{Lee:1984kr}.  As will be emphasized (Section 3) LFV requires a {\it source}, \ie, any operator that couples lepton generations.  For example, it is compelling that supersymmetric models with a neutrino seesaw mechanism, given reasonable parameters consistent with experiment, simultaneously provide an explanation of why $\mu\to e\gamma$ has not yet been observed and a prediction that $\mu\to e\gamma$ may be observable with current technology.  LFV decays for this particular model was first calculated in ref.~\cite{Borzumati:1986qx} and comprehensively calculated in ref.~\cite{Hisano:1995cp}.  The MEG experiment is a dedicated $\mu^+ \to e^+ \gamma$ experiment and has been acquiring data since 2009 \cite{Adam:2009ci}.   Thus far, the null result of MEG has provided a 90\% confidence level upper limit on the $\mu^+ \to e^+ \gamma$ branching ratio of $5.7\times10^{-13}$.  Consider the flavor violation to be manifested in a 1-loop diagram with a chargino and sneutrino with a mass-insertion $(m^2_{\tilde L})_{12}$.  Then using an estimate from naive dimensional analysis
\beq
BR(\mu\to e\gamma) \sim \alpha^3 \frac{([m^2_{\tilde L}]_{12})^2}{G_F^2 m_S^8},
\eeq
\noindent the MEG result implies that the off-diagonal entry $(m^2_{\tilde L})_{12}$ of the left-handed slepton mass matrix should not be larger than roughly 0.06\% of the slepton masses.  Here, $m_S$ is the typical sparticle mass, assumed to be 150 GeV to accommodate an MSSM interpretation of the muon anomalous magnetic moment measurement.  This stringent constraint highlights the need for tools to rapidly calculate LFV observables.

\begin{table}[t]
\small
\centering
\begin{tabular}{lll}
\hline
Observable & Limit & Future\\
\hline
$\mu^+\to e^+\gamma$ & $5.7\times 10^{-13}$ & $10^{-13}$ MEG \cite{Adam:2009ci}\\
$\tau^+\to e^+\gamma$ & $3.3\times 10^{-8}$ & $2.3\times 10^{-9}$ SuperB \cite{superB}\\
$\tau^+\to \mu^+\gamma$ & $4.4\times 10^{-8}$  & $3\times 10^{-9}$ Belle II \cite{Abe:2010sj}, $1.8\times 10^{-9}$ \cite{superB}\\
$\mu\to eee$ & $1.0\times 10^{-12}$ & $10^{-15}$ MUSIC \cite{Yoshida:2009zz}, $10^{-16}$ Mu3e \cite{mu3e}\\
$\tau\to eee$ & $2.7\times 10^{-8}$ & $2\times 10^{-10}$ \cite{superB}\\
$\tau\to \mu\mu\mu$ & $2.1\times 10^{-8} $ & $1\times 10^{-9}$ \cite{Abe:2010sj}, $2\times 10^{-10}$ \cite{superB}\\
$\mu^-\,{\rm SiC} \to e^-\,{\rm SiC}$ & none & $10^{-14}$ DeeMe \\ 
$\mu^-\,{\rm Al} \to e^-\,{\rm Al}$ & none & $10^{-16}$ COMET \cite{comet}, Mu2e \cite{mu2e} \\
 $\mu^-\,{\rm Ti} \to e^-\,{\rm Ti}$ & $4.3\times 10^{-12}$ & $10^{-18}$ PRISM/PRIME \cite{Barlow:2011zz} \\
\hline
\end{tabular}
\caption{{\bf Rates calculated by \superlfv.}  Current experimental limits are listed at the 90\% confidence level \cite{pdg}.}
\label{observables}
\end{table}

MEG's $\mu\to e\gamma$ search is just one of many near future and potential experiments.  $\tau\to e\gamma$ and $\tau\to \mu\gamma$ will be probed at the LHC and future $B$-factories \cite{Abe:2010sj, superB} .  At minimum, these experiments should probe another order of magnitude in the various branching ratios; see Table \ref{observables}.  Additionally, there are two other classes of LFV experiments --- namely, the $e_i\to 3e_j$ and $\mu^-N \to e^-N$ processes.  The MUSIC project is a high-intensity muon source currently being constructed at Osaka University (Japan) and has proposed to extend the $\mu\to3e$ reach by 3 orders of magnitude \cite{Yoshida:2009zz}.  Also, the PSI (Switzerland) may upgrade its existing muon source that provides $10^8$ muons/sec to $10^9$ muons/sec.  If so, the proposed $\mu3e$ experiment may improve the $\mu\to3e$ reach by 4 orders \cite{mu3e}.  The tau variants, $\tau\to3e$ and $\tau\to3\mu$ will also be further probed by the LHC and future $B$-factories.

Another class of LFV experiments is the search for neutrinoless decays of ground state muonic atoms, referred to as $\mu-e$ conversion in atomic nuclei (``muon conversion,'' for brevity).  By expectations from the SM, when the muon decays, either a $W^-$ boson is emitted outwards producing ordinary three-body muon decay or the $W^-$ is captured by the nucleus, lowering the proton count by one (``muon capture'').  Muon conversion seeks neutrinoless muon decay, with a free electron emitted.  {\it Coherent} muon conversion refers to interactions that involve all nucleons, hence the amplitude scales with the number of nucleons.  Coherent muon conversion requires an unexcited nucleus in the final state.  There are four potential future muon conversion experiments, DeeMe, COMET, Mu2e, and PRISM/PRIME \cite{deeme, mu2e, comet, Barlow:2011zz}.  Each target atom provides a unique observable, as interactions vary for different nuclei.

\superlfv calculates the amplitudes of observables listed in Table \ref{observables} at one-loop level.  {\tt SuperLFV} accepts an SLHA spectrum file, which contains the couplings and mass parameters of the MSSM sparticles, and outputs the LFV rates.  While this functionality is also included in related tools, {\tt SPheno}\cite{spheno}, {\tt SuSeFLAV} \cite{Chowdhury:2011zr}, and partially by {\tt SUSY\_FLAVOR} \cite{susy_flavor}, those tools currently do not accept externally generated SLHA spectrum files.  \superlfv is intended to be universal in the sense that it may be used with any tool that generates a SLHA spectrum file.

This separation of the spectrum generator and the observables calculator has the following merits.  The source of LFV is allowed to be arbitrary.  To date, all available SLHA-compliant spectrum generators use the supersymmetric seesaw mechanism as the source of LFV.  However, grand unified theories \cite{Hall:1985dx} and flavor violating $D$-terms from decoupled $Z'$ models \cite{Murakami:2001hk} are examples of other LFV sources that do not introduce new low scale particles.  The SLHA spectrum file that \superlfv uses as inputs may be generated by code or hand.  Also, this approach facilitates redundant checks of existing tools and allows focus on LFV-specific future enhancements to \superlfv.  Lastly, using an SLHA spectrum calculator, the SLHA allows one to specify arbitrary boundary conditions at a high scale (Planck, GUT, etc.) for an arbitrary model.  The spectrum calculator will then evolve all parameters to the low scale, where \superlfv may compute the LFV observables.

Another feature of \superlfv is the option to report all rates in terms of  classes of contributing diagrams -- \ie, contributions to $\mu\to e\gamma$ from diagrams involving charginos vs.~neutralinos may be reported separately, as well as their interference contribution.  Furthermore, any relevant coupling constants, effective and tree-level, are also optionally reported with a similar accounting for various contributions.  Hence, \superlfv is also designed to provide insight to the analytical behavior of the dynamics within the various LFV rates.

This article is organized as follows.  Relevant conventions are defined in Section 2.  The physics of the included observables and their explicit expressions are reproduced in Section 3.  In Section 4, the internal calculations of the code are detailed.  All approximations used are disclosed in Sections 3 and 4 and Appendix A.  The installation and operation of \superlfv is documented in Section 5.

\section{Lagrangian and conventions}

All sign conventions, normalizations, conjugations, and matrix definitions are made to match the SLHA conventions, though the actual notation used may differ superficially in symbol choice.  Irrespective of the high scale sources that generate the input parameters, \superlfv assumes only the MSSM fields, including three Majorana left-handed neutrinos, as its complete effective theory.  For clarity, the relevant notation used throughout this article is defined here.

The superpotential is written
\beq
W \equiv \epsilon_{\alpha\beta}(-y_u^{ij} \, H_u^\alpha \, Q_i^\beta \, \bar u_j
+ y_d^{ij} \, H_d^\alpha \, Q_i^\beta \, \bar d_j
+ y_e^{ij} \, H_d^\alpha \, L_i^\beta \, \bar e_j
+ \mu H_u^\alpha  H_d^\beta) .
\eeq
\noindent
where $\epsilon_{12} \equiv +1$.  Superfield notation is suppressed.  Per the SLHA, the quark and lepton multiplets are written in the super-CKM and super-PMNS bases, in which quarks, charged leptons, and neutrinos are written as their mass states.  That is, the sfermion components of the superfields have undergone the same rotations as their fermion counterparts.  The Higgs vevs are defined as $\langle H^0_u \rangle \equiv v_u / \sqrt2$ and $\langle H^0_d \rangle \equiv v_d / \sqrt2$, such that $v_u^2 + v_d^2 = (246\,{\rm GeV})^2$.  Accordingly, Yukawa couplings are defined at tree-level with $\sqrt 2$ factors; \eg, $m_u \equiv y_u v_u/\sqrt2$.  The gauge couplings of the hypercharge, weak, and strong interactions are denoted $g_Y$, $g_2$, and $g_3$.  In all that follows, it is assumed that all couplings and mass parameters are evaluated at the supersymmetry breaking scale (``soft scale,'' for brevity); \eg, $m_{\rm soft} = \sqrt {m_{\tilde t_1} m_{\tilde t_2}}$ or any arbitrarily defined scale.

The SLHA requires that all running couplings and masses are defined in the $\overline{\rm DR}$ scheme.  There are some exceptions to this rule in defining the {\tt SMINPUTS} block.  However, the non-$\overline{\rm DR}$ parameters of {\tt SMINPUTS} are only used by spectrum calculators.  As an observables calculator, only the output of those spectrum calculators are relevant to {\tt Super LFV}.  \eg, A spectrum calculator requires $\alpha_{\rm em}^{-1}(m_Z)$ defined via $\overline{\rm MS}$ in the {\tt SMINPUTS} block, but the spectrum calculator is required to output gauge couplings in the {\tt GAUGE} block in the $\overline{\rm DR}$ scheme, which is then used by \superlfv.  Hence, all \superlfv running parameters are defined via $\overline{\rm DR}$.  

The soft parameters are written
\bea
-\lag_{\rm soft} &\equiv& \left( \frac 1 2 M_1 \bino \bino + \frac 1 2 M_2 \wino \wino + \frac 1 2 M_3 \tilde g \tilde g + {\rm h.c.} \right) \nonumber \\
&& +\,\, \tilde Q^\dagger m^2_{\tilde Q} \tilde Q
+ \tilde u_{R}^t m^2_{\tilde u} \tilde u_R^*
+ \tilde d_{R}^t m^2_{\tilde d} \tilde d_R^*
+ \tilde L^\dagger m^2_{\tilde L} \tilde L
+ \tilde e_{R}^t m^2_{\tilde e} \tilde e_R^* \nonumber\\
&& +\,\, (-H_u \tilde Q a_u \tilde u_{R}^*
+ H_d \tilde Q a_d \tilde d_{R}^*
+ H_d \tilde L a_e \tilde e_{R}^*
+ {\rm h.c.}) \nonumber\\
&& +\,\, m^2_{H_u} H_u^* H_u + m^2_{H_d} H_d^* H_d
+ (b H_u H_d + {\rm h.c.})
\eea
All fermions are written in 2-component Weyl notation, unless specified.  Flavor indices are suppressed, though symbols are arranged in matrix multiplication order; \ie, $\tilde Q^\dagger m^2_{\tilde Q} \tilde Q \equiv \tilde Q_i^\dagger (m^2_{\tilde Q})_{ij} \tilde Q_j$  Isospin indices have been suppressed.  The trilinear couplings follow the same isospin conventions as their Yukawa counterparts; \ie, $H_u \tilde Q a_u \tilde u_{R}^* \equiv  H_u^\alpha \epsilon_{\alpha\beta} \tilde Q^\beta a_u \tilde u_{R}^*$.  Higgs soft mass operators have suppressed the following suppressed indices:  $H_u^*H_u = H_{u\alpha}^*H_u^\alpha$, $H_d^*H_d = H_{d\alpha}^*H_d^\alpha$, and $H_uH_d = \epsilon_{\alpha\beta} H_u^\alpha H_d^\beta$.

For the calculation of observables, all couplings are understood to be in the mass basis, which require mixing parameters.  The remainder of this section will define the conventions used in determining masses and mixings. Starting with neutralinos, the mass operators are written as
\beq
-\lag_{{\tilde\chi^0}} \equiv \frac 1 2
[\bino \,\, \wino^3 \,\, \higgsino_d^0 \,\, \higgsino_u^0]
\left[ \begin{array}{cccc}
M_1 & 0 & -g_Y v_d / 2 & g_Y v_u / 2 \\
0 & M_2 & g_2 v_d / 2 & -g_2 v_u / 2 \\
-g_Y v_d / 2 & g_2 v_d / 2 & 0 & -\mu \\
g_Y v_u / 2 & -g_2 v_u / 2 & -\mu & 0
\end{array}
\right]
\left[ \begin{array}{c}
\bino \\ \wino^3 \\ \higgsino_d^0 \\ \higgsino_u^0
\end{array} \right] + {\rm h.c.}
\eeq
Gaugino spinors have been redefined by the convention $-i\tilde A \to \tilde A$, such that the off-diagonal $2\times2$ sub-matrices are real.  The unitary transformation matrix $N$ that yields the neutralino mass states $\tilde\chi^0_A$  is defined by $\tilde\chi^0_A = N_{AB} \lambda_B$ where $\lambda \equiv [\tilde B,\, \tilde W^3,\, \tilde H_d^0,\, \tilde H_u^0]$.  Throughout this article, capital indices $A$ and $B$ denote neutralino or chargino indices.  By the SLHA, all sparticle mass indices are ordered by absolute value, with index 1 representing the lightest state.  The SLHA allows for signed (real) masses for neutralinos and charginos.

Chargino mass operators are written
\bea
-\lag_{{\tilde\chi^\pm}} &=&
[\wino^- \,\, \higgsino_u^-]
\left[
\begin{array}{cc}
M_2 & g_2 v_u/\sqrt 2\\
g_2 v_d/\sqrt 2 & \mu
\end{array}
\right]
\left[
\begin{array}{c}
\wino^+ \\
\higgsino_d^+
\end{array}
\right]  + {\rm h.c.} \nonumber\\
&\equiv& \lambda^- M_{\tilde\chi^+} \lambda^+
\label{charginoMassMatrix}
\eea
The unitary transformation matrices $U$ and $V$ that yield the chargino mass states $\tilde\chi^\pm_i$ are defined by $\chi_i^- \equiv U_{ij}\lambda^-_j$ and $\chi^+_i \equiv V_{ij}\lambda^+_j$.

To obtain the masses and mixings of sfermions in their mass states, it is conventional to form $6\times6$ mass matrices for charged sleptons, up squarks, and down squarks and a $3\times3$ mass matrix for left-handed sneutrinos; \ie,
\beq
-\lag_{\tilde f} \equiv
[\tilde u_L^\dagger \tilde u_R^\dagger]
M^2_{\tilde u}
\left[ \begin{array}{c}
\tilde u_L \\ \tilde u_R
\end{array} \right]
+
[\tilde d_L^\dagger \tilde d_R^\dagger]
M^2_{\tilde d}
\left[ \begin{array}{c}
\tilde d_L \\ \tilde d_R
\end{array} \right]
+
[\tilde e_L^\dagger \tilde e_R^\dagger]
M^2_{\tilde e}
\left[ \begin{array}{c}
\tilde e_L \\ \tilde e_R
\end{array} \right]
+
\tilde \nu_L^\dagger M^2_{\tilde\nu} \tilde \nu_L .
\eeq
Flavor indices have been suppressed, though $\tilde u_L \equiv [\tilde u_L, \tilde c_L, \tilde t_L]$, for example.  Each $6\times6$ mass matrix of sfermion type $\tilde f$ has the form
\beq
M^2_{\tilde f} \equiv
\left[ \begin{array}{cc}
m^2_{\tilde f LL} & m^{2\dagger}_{\tilde f LR} \\
m^2_{\tilde f LR} & m^2_{\tilde f RR}
\end{array} \right].
\eeq

For squarks,
\bea
m^2_{\tilde u LL} &=& V_{\rm CKM} m^2_{\tilde Q} V_{\rm CKM}^\dagger+ \frac 1 2 |y_u v_u|^2 + D_{u_L} , \\
m^2_{\tilde u RR} &=& m^{2t}_{\tilde u} + \frac 1 2 |y_u v_u|^2 + D_{u_R} , \\
m^2_{\tilde u LR} &=& \frac 1 {\sqrt 2}(a_u^t v_u - y_u^t v_d \mu^*) ,\\
m^2_{\tilde d LL} &=& m^2_{\tilde Q} + \frac 1 2 |y_d v_d|^2 + D_{d_L} , \\
m^2_{\tilde d RR} &=& m^{2t}_{\tilde d} + \frac 1 2 |y_d v_d|^2 + D_{d_R} , \\
m^2_{\tilde d LR} &=& \frac 1 {\sqrt 2}(a_d^t v_d - y_d^t v_u \mu^*) .
\eea
Again, flavor indices have been suppressed.  For Yukawa matrices, $|y_u|^2 = y_u^\dagger y_u$, despite being defined to be real and diagonal.  The $D$-term for sfermion type $\tilde f$ is
\beq
D_{\tilde f} = \frac 1 4 (g_Y^2 Y - g_2^2 t^3) (v_u^2 - v_d^2) {\bf 1}
\eeq
where $Y$ is the hypercharge generator and $t^3$ is the third SU(2) generator.  Both generators are to be replaced by the corresponding eigenvalues of the fields they act on.  {\bf 1} is the $3\times3$ identity matrix.  Hypercharge assignments are defined via $q = t^3 + Y$, where $q$ is the electric charge of the representation of the sfermion.  That is, $q$ for left-handed representations of right-handed multiplets are opposite the charge of its Direct fermion component.  The generator $t^3$ is defined as half the Pauli spin matrix $\sigma^3$.

For charged sleptons,
\bea
m^2_{\tilde e LL} &=& m^2_{\tilde L} + \frac 1 2 |y_e v_d|^2 + D_{e_L} , \\
m^2_{\tilde e RR} &=& m^{2t}_{\tilde e} + \frac 1 2 |y_e v_d|^2 + D_{e_R} , \\
m^2_{\tilde e LR} &=& \frac 1 {\sqrt 2}(a_e^t v_d - y_e^t v_u \mu^*) .
\eea

Except for sneutrinos, sfermions of type $\tilde f$ are rotated via a $6\times6$ matrix $R_{\tilde f}$; \ie, 
\beq
\tilde f_I = (R_{\tilde f})_{IJ}
\left[ \begin{array}{c}
\tilde f_L \\
\tilde f_R
\end{array}
\right]_J ,
\eeq
where
\beq
R_{\tilde f} M^2_{\tilde f} R_{\tilde f}^\dagger
\eeq
yields a diagonal matrix with the mass eigenvalues as the diagonal entries.

The sneutrino mass matrix is written
\beq
M^2_{\tilde \nu} = 
U_{\rm PMNS}^\dagger m^2_{\tilde L} U_{\rm PMNS} + D_{\nu_L} .
\label{snuMasses}
\eeq
This expression assumes that $m^2_{\tilde L}$ is in the super-PMNS basis in which both fermionic and scalar components of the charged lepton superfield are both rotated together, such that charged leptons are in their mass states.  \ie, $e_L \to V_e e_L$ and $e_R \to U_e e_R$ with the same done to the scalar components.  In the super-PMNS basis, the left-handed sneutrino states are rotated using the same transformation that rotates left-handed neutrinos to their mass states.  \ie, $\nu_L \to V_\nu \nu_L$ and $\tilde \nu_L \to V_\nu \tilde \nu_L$.  Through the weak interaction operator $W^-_\mu[\bar e_{Li} \gamma^\mu \nu_{Li}]$, the PMNS matrix is defined as $U_{\rm PMNS} \equiv V_e^\dagger V_\nu$.

\section{Branching ratios}

In this section, the physics and expressions for the included LFV rates are reviewed.  For each process, expectations for analytical behavior is discussed, as well as any approximations or omitted amplitudes in the \superlfv code.

LFV requires a source.  Naturally, the LFV source will control the analytical behavior of LFV observables.  Mainstream supersymmetry breaking models are generally designed to be free of observable charged LFV.  This is due to the flavor problem of supersymmetry and the assumed irrelevance of flavor violation in probing the origin of electroweak symmetry breaking.  The source is typically heavy particles, though light states that, say, couple only to third generation leptons are also conceivable.

We now commit to the MSSM in all that follows.  High scale dynamics may generate off-diagonal soft masses in the left-handed slepton mass matrix $m^2_{\tilde L}$, the right-handed slepton mass matrix $m^2_{\tilde R}$, or the matrix of trilinear couplings $a_e$.  This allows the mass states of leptons and sleptons to exhibit LFV at tree-level interactions with neutralinos and charginos.  Below the soft scale, the primitive LFV couplings will generate effective operators that only involve SM particles.  The primitive couplings and any effective operators they generate are listed in Appendix A.

The mainstream sources of LFV are neutrino seesaw models and grand unified theories (GUTs).  The right-handed neutrinos of seesaw models will radiatively generate off-diagonal entries to the left-handed slepton mass matrix $m^2_{\tilde L}$, due to Yukawa interactions with left-handed leptons.  In this scenario, contributions to off-diagonal entries of the right-handed slepton mass matrix $m^2_{\tilde e}$ will be loop-suppressed.

For SU(5) GUT models (with no seesaw mechanism), quark-mixing for the unified quark-lepton multiplets necessarily means lepton-mixing, induced by RG running between the Planck scale and GUT scale.  When the SU(5) gauge group is spontaneously broken at the GUT scale, the MSSM as an effective model will be generated with off-diagonal entries to $m^2_{\tilde e}$.  Hence, $m^2_{\tilde e}$ will be the dominant LFV source.  These examples demonstrate that the LFV source is a model-building choice.  While neither class of mainstream models yield the trilinear couplings $(H_d \tilde L a_e \tilde e_{R}^* + {\rm h.c.})$ as the dominant LFV source, it remains a possibility.

In the following calculations, we follow the comprehensive work of ref.~\cite{Hisano:1995cp} with corrections to the $e_i\to3e_j$ decay provided by ref.~\cite{Arganda:2005ji}.

\subsection{$e_i \to e_j \gamma$}

The effective lagrangian for  $e_i \to e_j \gamma^*$, defined at the scale of the lepton mass $m_{e_i}$, is
\bea
-\lag &\supset& e q^2 A_\mu \bar e_i \gamma^\mu (A_{1L}^{ij} P_L + A_{1R}^{ij} P_R) e_j \nonumber\\
&&+ \frac {em_{e_i}}{2} \bar e_i \sigma_{\mu\nu} F^{\mu\nu} (A_{2L}^{ij} P_L + A_{2R}^{ij} P_R) e_j + {\rm h.c.}
\label{emOperators}
\eea
The electromagnetic form factors $A_{1L}^{ij}$ and $A_{1R}^{ij}$ are defined to exclude the tree-level value $-\delta^{ij}$, and do not contribute to on-shell $e_i \to e_j \gamma$, as made explicit by factoring out the photon's external momentum squared $q^2$ from the couplings.  $A_{1L}$ and $A_{1R}$  will be relevant later to $e_i \to 3e_j$ and muon conversion, where off-shell photons contribute.

The $e_i \to e_j \gamma$ branching ratio definition deviates from the conventional definition of a branching ratio in which a partial width is compared to the total width.  Comparison of the partial width $\Gamma(e_i \to e_j \gamma)$ to $\Gamma(e_i\to e_j \bar\nu_j\nu_i)$ is the standard observable, evaluated in the limit $m_{e_j} \ll m_{e_i} \ll m_W$.  For $\tau$ decays, this definition deviates significantly from the conventional definition.  To lowest order, the partial width of $e_i\to e_j \bar\nu_j\nu_i$ is
\beq
\Gamma(e_i\to e_j \bar\nu_j\nu_i) = \frac {G_F^2 m_{e_i}^5}{192\pi^3}.
\eeq
Consequently, the effective theory given by eqn.~(\ref{emOperators}) yields
\beq
{\rm BR}(e_i \to e_j \gamma) = 48\pi^3 \frac{\alpha}{G_F^2}
\left(|A_{2L}^{ij}|^2 + |A_{2R}^{ij}|^2 \right).
\eeq

The following remarks are regarding the expected analytical behavior of $e_i \to e_j \gamma$.

\bi

\item At 1-loop order, the couplings $A_{2L}^{ij}$ and $A_{2R}^{ij}$ are generated by loops involving either charginos or neutralinos.  These form factors generally scale linearly $\tan\beta$.  This can be understood by considering diagrams involving all sparticles in their gauge-interaction states (``gauge states,'' for brevity).  Then diagrams can be distinguished as those with only gauginos in the loops {\it vs.}~those with higgsino-gaugino mixing in the loops.  Then the higgsino-gaugino mixing manifests as mass-insertion involving the up-type Higgs vev $v_u$.  Also, the Yukawa coupling of a lepton and down-Higgsino $H_d$ will contain $1/v_d$.  Together, the $\tan\beta$ dependence is formed.

\item Experimental constraints require off-diagonal soft masses of charged sleptons to be small.  Yet, off-diagonal masses for sneutrinos, in the Super-PMNS basis of eqn.~(\ref{snuMasses}), may be large.  Therefore, diagrams involving sneutrino mixings should dominate over those with slepton mixing.  Alternatively, chargino diagrams should generally dominate over neutralino diagrams.

\item The tensor operators of eqn.~(\ref{emOperators}) force a helicity change.  Amplitudes that change the lepton helicity from right to left will involve the Yukawa coupling of the decaying (heavier) lepton, for diagrams with higgsino-gaugino mixing.  This will dominate over diagrams that involve Yukawa coupling of the lighter final-state lepton.  If the LFV source is $m^2_{\tilde L}$, the initial lepton must be right-handed for the diagram to be proportional to larger Yukawa coupling.  Hence, $e_{Ri} \to e_{Lj} \gamma$ should dominate over $e_{Li} \to e_{Rj} \gamma$ when $m^2_{\tilde L}$ is the LFV source.  Equivalently, the $A_{2L}^{ij}$ coupling should dominate over $A_{2R}^{ij}$.  These statements are reversed if the LFV source is $m^2_{\tilde e}$.

\item Furthermore, if $m^2_{\tilde e}$ is the dominant LFV source, there are no charged winos in the loop diagrams.  Only amplitudes with loops involving either a pure bino or a bino-Higgsino mixing remain.

\item At one-loop, a non-holomorphic coupling of leptons to the up-type Higgs $H_u^0$ generates the effective coupling of the form $\bar e_{Ri} y_{ij}z_{ij} H_u^{0*} e_{Lj}$, if an LFV source exist for $m^2_{\tilde L}$, $m^2_{\tilde R}$, or $a_e$ \cite{Babu:2002et}.  Here, $z_{ij}$ parameterizes the radiatively-generated coupling.  This up-type Higgs couplings yields two more mechanisms for mediating LFV.  Presently, neither have been included in \superlfv.
\bi
\item Using this effective coupling at one-loop, the neutral Higgs bosons $h^0$, $H^0$, and $A^0$ will mediate $e_i \to e_j \gamma$ \cite{Hisano:2010es}.  However, such diagrams are doubly Yukawa suppressed.  At two-loops, diagrams classified as Barr-Zee diagrams avoid this suppression and dominate.  It can be demonstrated that for reasonably light sparticles, this Higgs-mediated LFV is negligible.  It becomes dominant when the ratio of the sparticle mass scale to the mass of the pseudo-scalar Higgs $m_{\rm soft}/m_A$ grows beyond roughly 40-50.
\item At tree-level, the Higgs propagator from the non-holomorphic coupling vertex may be connected to another lepton or nucleon to yield $e_i\to 3e_j$ \cite{Babu:2002et} and muon conversion \cite{Kitano:2003wn}.  This is irrelevant to $e_i \to e_j \gamma$, but pointed out to avoid confusion, as both types of Higgs mediated LFV enter in  other LFV observables.
\ei

\item If trilinear couplings are the dominant LFV source, the dominant loop diagram (cast in mass states) will involve only a neutralino propagator and slepton propagator.  A chargino would create a left-handed sneutrino.  The trilinear coupling would  flip the neutrino helicity to the decoupled right-handed sneutrino.  Hence, charginos do not contribute to this process.  The neutralino-slepton loop will be much like those with $m^2_{\tilde L}$ or $m^2_{\tilde e}$ mass-insertions.  In the gauge-interaction basis, there will be two one-loop amplitudes:  one with a pure bino and another with a bino-higgsino mixing.  Either can be made to dominate, resulting in non-trivial behavior.
\bi
\item In a loop with a pure bino for the neutralino propagator, there will be a single mass insertion by the trilinear coupling of $i v_d (a_e)_{ij} / \sqrt2$ along the slepton propagator.  The $\tan\beta$ dependence of this amplitude is controlled by the $v_d$ vev, or $v/\sqrt{1 + \tan^2\beta}$ .  The left-right mixing of the trilinear coupling is sufficient for the required helcity flip of the electromagnetic dipole operator.

\item In a loop with a bino-higgsino transition along the neutralino propagator, since the higgsino will cause one helicity flip, two left-right mass insertions are required along the slepton propagator.  One trilinear should be the LFV mass-insertion $i v_d (a_e)_{ij} / \sqrt2$.  The other mass-insertion should be the usual flavor-preserving left-right slepton mixing.  Hence, these three helicity flips satisfy the electromagnetic dipole operator.

Each left-right mixing introduces one factor of the Higgs vev $v_d$ to the amplitude.  The bino-higgsino transition inlcudes a $v_u$ factor, while the lepton-slepton-higgsino Yukawa coupling includes a $1/v_d$ factor.  This results in an overall $v_u v_d$ factor, or $v^2 \tan\beta/(1+ \tan^2\beta)$.  That is, this amplitude decreases with $\tan\beta$.  Also note that, if the flavor-preserving trilnear couplings are negligible, this amplitude itself becomes neglibible.
\ei
Unlike the $m^2_{\tilde L}$ and $m^2_{\tilde e}$ LFV sources, a LFV trilinear does not force the initial lepton to be left- or right-handed.  Therefore, the amplitudes for $e_{Ri} \to e_{Lj} \gamma$ and $e_{Li} \to e_{Rj} \gamma$ (or the magnitudes for $A_{2R}^{ij}$ and $A_{2L}^{ij}$) are expected to be comparable for trilinear LFV sources.

\item Loops with purely SM particles, \ie~$W$ bosons and neutrinos, are omitted in the \superlfv code, as it is for all other rates to follow.  In principle, neutrino oscillations will yield non-zero LFV.  However, the smallness of the neutrino masses place contributions to the $\mu\to e\gamma$ branching ratio on the order ${\mathcal O}(m_\nu/m_W)^4$, several orders below $10^{-50}$ for reasonable neutrino parameters.

\ei

\subsection{$e^-_i\to e^-_j e^-_j e^+_j$}

The amplitude for $e_i \to 3e_j$ decay includes diagrams with photon exchange, $Z$ exchange, Higgs exchange, and neutralino-slepton (chargino-sneutrino) boxes.  Defined at the scale of the lepton mass $m_{e_i}$, the effective lagrangian for $e_i \to 3e_j$ decay that provides photon exchange is
\beq
-\lag \supset ie^2 [\bar e_j \gamma_\mu (A_{1L}^{ij} P_L + A_{1R}^{ij} P_R) e_i][\bar e_j \gamma^\mu e_j]
+ \frac {e^2 m_{e_i} }{q^2} [\bar e_j \sigma_{\mu\nu} q^\nu (A_{2L}^{ij} P_L + A_{2R}^{ij} P_R) e_i ] [\bar e_j \gamma^\mu e_j] .
\eeq
$Z$ exchange is governed by
\beq
-\lag \supset \frac {ig_2^2}{m^2_Z}[\bar e_j \gamma_\mu(F_L^{ij} P_L +F_R^{ij} P_R) e_i][\bar e_j \gamma^\mu(Z_{e_L} P_L + Z_{e_R} P_R) e_j]
\eeq
where $Z_\psi$ is the tree-level $Z$ coupling to a chiral fermion $\psi$, given by
\beq
Z_\psi = t^3_\psi - q_\psi \sin^2\theta_w .
\eeq
Neutralino-slepton (chargino-sneutrino) box diagrams are governed by
\bea
-\lag &\supset& ie^2 B_{1L}^{ij} [\bar e_j \gamma_\mu P_L e_i][\bar e_j \gamma^\mu P_L e_j]
+ ie^2 B_{1R}^{ij} [\bar e_j \gamma_\mu P_R e_i][\bar e_j \gamma^\mu P_R e_j] \nonumber\\
&&+ ie^2 B_{2L}^{ij} [\bar e_j \gamma_\mu P_L e_i][\bar e_j \gamma^\mu P_R e_j]
+ ie^2 B_{2R}^{ij} [\bar e_j \gamma_\mu P_R e_i][\bar e_j \gamma^\mu P_L e_j] \nonumber\\
&&+ ie^2 B_{3L}^{ij} [\bar e_j P_L e_i][\bar e_j P_L e_j]
+ ie^2 B_{3R}^{ij} [\bar e_j P_R e_i][\bar e_j P_R e_j] \nonumber\\
&&+ ie^2 B_{4L}^{ij} [\bar e_j \sigma_{\mu\nu} P_L e_i][\bar e_j \sigma^{\mu\nu} P_L e_j]
+ ie^2 B_{4R}^{ij} [\bar e_j \sigma_{\mu\nu} P_R e_i][\bar e_j \sigma^{\mu\nu} P_R e_j].
\eea
All form factors are supplied in Appendix A.

As with $e_i\to e_j\gamma$, the $e_i\to 3e_j$ branching ratio is defined as the ratio of its partial width to the $e_i\to e_j \bar\nu_j\nu_i$.  This results in a branching ratio of \cite{Hisano:1995cp, Arganda:2005ji}
\bea
{\rm BR}(e^-_i\to e^-_j e^-_j e^+_j) &=& \frac{6\pi^2\alpha^2}{G_F^2} [
|A_{1L}^{ij}|^2 + |A_{1R}^{ij}|^2
- 4\re(A_{1L}^{ij} A_{2R}^{ij*} + A_{2L}^{ij} A_{1R}^{ij*})
\nonumber\\
&&+ \,\, (|A_{2L}^{ij}|^2 + |A_{2R}^{ij}|^2) \left(\frac {16}3\ln\frac{m_{e_i}}{m_{e_j}} - \frac{22}3 \right) \nonumber\\
&&+ \,\, \frac 16 (|B_{1L}^{ij}|^2 + |B_{1R}^{ij}|^2) + \frac 13 (|B_{2L}^{ij}|^2 + |B_{2R}^{ij}|^2) \nonumber\\
&&+ \,\, \frac 1{24} (|B_{3L}^{ij}|^2 + |B_{3R}^{ij}|^2) + 6 (|B_{4L}^{ij}|^2 + |B_{4R}^{ij}|^2) \nonumber\\
&&- \,\, \re(B_{3L}^{ij} B_{4L}^{ij*} + B_{3R}^{ij} B_{4R}^{ij*}) \nonumber\\
&&+ \,\, \frac 23 \re (A_{1L}^{ij} B_{1L}^{ij*} + A_{1R}^{ij} B_{1R}^{ij*} + A_{1L}^{ij} B_{2L}^{ij*} + A_{1R}^{ij} B_{2R}^{ij*}) \nonumber\\
&&- \,\, \frac 43 \re (A_{2R}^{ij} B_{1L}^{ij*} + A_{2L}^{ij} B_{1R}^{ij*} + A_{2L}^{ij} B_{2R}^{ij*} + A_{2R}^{ij} B_{2L}^{ij*}) \nonumber\\
&&+ \,\, \frac 13 \{ 2|F_{LL}^{ij}|^2 + 2|F_{RR}^{ij}|^2 + |F_{LR}^{ij}|^2 + |F_{RL}^{ij}|^2 \nonumber\\
&&+ \,\, 2 \re(B_{1L}^{ij} F_{LL}^{ij*} + B_{1R}^{ij} F_{RR}^{ij*} + B_{2L}^{ij} F_{LR}^{ij*} + B_{2R}^{ij} F_{RL}^{ij*}) \nonumber\\
&&+ \,\, 4 \re(A_{1L}^{ij} F_{LL}^{ij*} + A_{1R}^{ij} F_{RR}^{ij*}) + 2\re(A_{1L}^{ij} F_{LR}^{ij*} + A_{1R}^{ij} F_{RL}^{ij*}) \nonumber\\
&&- \,\, 8 \re(A_{2R}^{ij} F_{LL}^{ij*} + A_{2L}^{ij} F_{RR}^{ij*}) - 4 \re (A_{2L}^{ij} F_{RL}^{ij*} + A_{2R}^{ij} F_{LR}^{ij*})]
\eea
where
\bea
F_{LL}^{ij} &=& \frac {F_L^{ij} Z_{e_L}}{\sin^2\theta_w m^2_W} ,\\
F_{RR}^{ij} &=& F_{LL}^{ij}|_{L \leftrightarrow R} ,\\
F_{LR}^{ij} &=& \frac {F_L^{ij} Z_{e_R}}{\sin^2\theta_w m^2_W} ,\\
F_{RL}^{ij} &=& F_{LR}^{ij}|_{L \leftrightarrow R}.
\eea
The following is noted about expectations for the $e_i \to 3e_j$ branching ratio.

\bi

\item On-shell photon exchange dominates due to the $\tan^2\beta$ enhancement in the branching ratio and a relatively large logarithmic enhancement $ (\frac {16} 3 \ln\frac{m_{e_i}}{m_{e_j}} - \frac{22} 3)$ relative to all other couplings, which are all defined to be of dimension GeV${}^{-2}$ (for comparison with the Fermi constant $G_F$).  It can be verified with the \superlfv code that the photon-exchange diagrams dominates these decays, with all other couplings typically being two to several orders of magnitude smaller.  Even in the case of low $\tan\beta$, the logarithmic enhancement of the dipole exchange contribution is sufficient to ignore all other contributions to percentile accuracy.  This yields a quite rigid prediction of

\beq
\frac {{\rm BR}(e_i \to 3e_j)}{{\rm BR}(e_i \to e_j\gamma)} = \frac {2\alpha}{3\pi} \left( \ln\frac{m_{e_i}}{m_{e_j}} - \frac{11}{8} \right) .
\eeq
This provides numerical values of roughly 0.61\% ($\mu\to3e$), 0.22\% ($\tau\to3\mu$), 1.1\% ($\tau\to3e$).

\item The $Z$-exchange diagrams can be classified as eight types, four each for neutralino and chargino loops.  Within this division, two diagrams involve loops with a $Z$ attached to either the neutralino (chargino) or slepton (sneutrino).  The other two diagrams are LFV self-energy loops on either the initial lepton or the final lepton.  The analytical behavior of these $Z$-exchange diagrams was elucidated by ref.~\cite{Hirsch:2012ax}.  These authors divide contributions into those with couplings $F_R$ and $F_L$ (see Appendix A).  $F_R$ are all Yukawa suppressed diagrams.  Next, they demonstrate that, in the absence of chargino-mixing, all contributions by a pure charged wino cancel exactly within $F_L$.

\item As mentioned earlier, the non-holomorphic Higgs coupling is omitted in the photon exchange and Higgs exchange amplitudes for $e_i \to 3e_j$.  Despite a $\tan^6\beta$ scaling, Higgs-exchange is unable to overcome Yukawa suppression for both muon and tau decays.  In the most complete study of $e_i \to 3e_j$, it is demonstrated that, in the then-experimentally allowed parameter space of minimal supergravity (mSUGRA), Higgs exchange remains a few orders of magnitude smaller than photon exchange for $\tan\beta$ values as high as 50 \cite{Arganda:2005ji}.  Yet, under more extreme conditions of heavy sparticles, a light pseudo-scalar neutral Higgs $A^0$, and large $\tan\beta$, Higgs exchange may compete with photon exchange.

\ei

\subsection{Muon conversion $\mu^- N \to e^-N$}

This process is governed by the same classes of diagrams as $e_i \to 3e_j$.  For QCD concerns, only quarks lighter than the QCD scale are included the effective theory.  Furthermore, strange quarks do not contribute to coherent muon conversion for the following reasons.  In the limit of isospin symmetry amongst light quarks, there is no vectorial coupling to strange quarks in the nucleus.  The axial-vector couplings to strange quarks yields an incoherent interaction due to spin transitions and is therefore neglected.  At the quark level, the lagrangian for photon exchange is
\bea
-\lag &\supset& e^2 [\bar e_j \gamma_\mu (A_{1L}^{ij} P_L + A_{1R}^{ij} P_R) e_i] \cdot \sum_q q_q [\bar q \gamma^\mu q] \nonumber\\
&&- \frac {ie^2 m_{e_i} }{q^2} [\bar e_j \sigma_{\mu\nu} q^\nu (A_{2L}^{ij} P_L + A_{2R}^{ij} P_R) e_i ] \cdot \sum_q q_q [\bar q \gamma^\mu q]
\eea
where the sum $\Sigma_q$ is over up and down quarks with electric charge $q_q$.  $Z$ exchange is described by
\beq
\lag \supset \frac 12 \frac {g_Z^2}{m^2_Z}[\bar e \gamma_\mu (F_L^{ij} P_L + F_R^{ij} P_R) \mu]  \sum_q (Z_{q_L} + Z_{q_R}) [\bar q \gamma^\mu q] .
\eeq
The neutralino-slepton (chargino-sneutrino) box diagrams are described by
\beq
\lag \supset e^2 \sum_q [\bar e \gamma_\mu (D_{q_L} P_L + D_{q_R} P_R) \mu] [\bar q \gamma^\mu q] .
\eeq
The four-point effective couplings $F_L$, $F_R$, $D_{q_L}$, and $D_{q_R}$ are supplied in Appendix A.  Note that Higgs exchange diagrams are currently omitted.  To calculate muon conversion amplitudes, the quark-level effective lagrangian is then used to create an effective theory at the nucleon-level \cite{Bernabeu:1993ta}.

The muon conversion branching ratio is defined as the ratio of muon conversion partial width to the muon capture partial width, where the final state nucleus of muon capture may be excited. That is,
\beq
{\rm BR}(\mu^-N \to e^-N) \equiv \frac{\Gamma(\mu^-N \to e^-N)}{\Gamma(\mu^-N \to \nu_\mu N')} .
\eeq
The {\it coherent} muon conversion decay width for a muonic atom of atomic number $Z$ and $N$ nucleons is \cite{Hisano:1995cp}
\bea
\Gamma(\mu^-N \to e^-N) &=& 4\alpha^5 \frac {Z_{\rm eff}^4} Z |F_N|^2 m_\mu^5 [|Z(A_{1L}^{21} - A_{2R}^{21}) - (2Z + N) \bar D_{u_L} - (Z+2N) \bar D_{d_L}|^2 \nonumber\\
&& + |Z(A_{1R}^{21} - A_{2L}^{21}) - (2Z + N) \bar D_{u_R} - (Z+2N) \bar D_{d_R}|^2]
\label{muonConversionWidth}
\eea
where the $Z$-exchange and box contributions have been combined as
\bea
\bar D_{q_L} &=& D_{q_L} + \frac {(Z_{q_L} + Z_{q_R}) F_L^{21}} {2\sin^2\theta_w m^2_W} , \\
\bar D_{q_L} &=& \bar D_{q_L}|_{L \leftrightarrow R}.
\eea
$Z_{\rm eff}$ and $|F_N|$ describe an effective nuclear charge and nuclear matrix element, respectively, and are defined in the pioneering framework provided by Weinberg and Feinberg \cite{Weinberg:1959zz}.  Table \ref{conversionTargets} lists the values used in the code.

\begin{table}[t]
\small
\centering
\begin{tabular}{lllll}
& $Z_{\rm eff}$ & $|F_N|$ & $\Gamma(\mu N\to\nu_\mu N')$ & $R$\\
\hline\\[-2mm]
${}_{13}^{27}$Al & 11.62 & 0.64 & $0.7054\times10^6\,{\rm s}^{-1}$ & 351\\
${}_{22}^{48}$Ti & 17.61 & 0.535 & $2.59\times10^6\,{\rm s}^{-1}$ & 203\\
\hline
\end{tabular}
\caption{{\bf Nuclear data for muon conversion.}  Here, $Z_{\rm eff}$ and $|F_N|$ are the effective nuclear charge and a nuclear matrix element that describes charge distribution, respectively, of eqn.~(\ref{muonConversionWidth}).  Estimates for  $Z_{\rm eff}$ and $|F_N|$ are from \cite{Chiang:1993xz}.  Estimates for the experimental muon capture partial widths $\Gamma(\mu N\to\nu_\mu N')$ are from \cite{Suzuki:1987jf}}.
\label{conversionTargets}
\end{table}

To address the nuclear aspects, a few assumptions are made.  Currently, \superlfv calculates muon conversion using the general approach of Weinberg and Feinberg, applied to the MSSM by Hisano, \etal~\cite{Hisano:1995cp}.  These assumptions are listed.
\begin{enumerate}
\item For $Z$-exchange and box diagrams, only vector currents are expected to be relevant, since axial-vector quark couplings yield spin-dependent nuclear effects.  This approximation is valid for sufficiently heavy isotopes.
\item The densities of protons and neutrons are assumed to be equal and constant.
\item The muon is treated non-relativistically, and the muon wavefunction overlap with the nucleus is taken to be a constant.
\item $F_N$ is a weighted overlap of the muon wavefunction and the nucleons, and hence applies the approximations of the previous two points.
\item The final state electron is treated non-relativistically as a plane wave, undistorted by the nuclear electric charge.
\item The electron mass is neglected.
\end{enumerate}

The state-of-the-art technique of addressing the nuclear aspects is currently authored by Kitano, \etal~\cite{Kitano:2003wn}.  Those authors apply the approach of Czarnecki, \etal~\cite{Czarnecki:1998iz}, which includes an improved scheme for nuclear considerations of off-shell photon exchange.  These groups also solved for the initial-state muon and final-state electron wavefunctions via the Dirac equation.  Kitano, \etal~also applied various schemes for nucleon distributions and experimental neutron distributions obtained from pionic atoms.  They find that coherent muon conversion rates for heavy elements can differ by $\sim 20-30$\% from the method of \cite{Bernabeu:1993ta}.

The following points are noted about expectations for the muon conversion rate.

\bi

\item For much of the MSSM parameter space, the photon-exchange amplitudes dominate the muon conversion rate.  In this case, there is roughly a fixed correlation of order $\alpha$ between a muon conversion rate and BR($\mu\to e\gamma$) that only depends on the atomic element.  For example, using the framework of Weinberg and Feinberg and the nuclear approximation scheme of Bernabeu, \etal, one arrives at a relation
\beq
\frac {BR(\mu\to e\gamma)}{\rm BR(\mu^- \, {\rm N} \to e^- \, {\rm N})} \approx R
\eeq
where values for $R$ are listed in Table \ref{conversionTargets}. This convenient guideline breaks down in the limits of both small and large $\tan\beta$, as noted below.

\item Unlike $e_i \to 3e_j$, on-shell photon exchange is not boosted by a large logarithm.  For small $\tan\beta$ and $\mu<0$, the on-shell photon exchange contribution may have large cancellations amongst its gauge state amplitudes \cite{Hisano:1995cp}.

\item Also unlike $e_i \to 3e_j$, Higgs exchange is not doubly Yukawa suppressed.  The neutral Higgs bosons couple proportional to the nucleon mass, via the superconformal anomaly.  For this reason and the previous point, Higgs exchange may be demonstrated to dominate for experimentally accessible parameter space.  For example, for mSUGRA parameters of unified scalar and gaugino soft masses of 1 TeV at the unification scale, \ie~$m_0 = m_{1/2} =$ 1 TeV, muon conversion for aluminum begins to deviate from photon exchange dominance for a heavy Higgs lighter than 400 GeV \cite{Kitano:2003wn}.

\ei

\section{Numerical calculations for the LFV observables}

In this section, the procedure of the numerical calculations is made transparent.

\subsection{Calculation procedure}

An SLHA spectrum file supplies all input for the calculation of amplitudes.  For branching ratios evaluated at low energy, couplings and masses that arise due to solely to kinematics are evaluated using hard-coded values compiled by the Particle Data Group (PDG); \eg, $\alpha(0)$ or pole masses.  The values of all Yukawa and gauge couplings and running mass parameters are read in at the scale supplied by the SLHA ``{\tt Q=}'' convention. For example, immediately following the header {\tt BLOCK GAUGE Q= 1.00000000E+03}, gauge couplings are supplied at the RG scale of 1 TeV.  These running parameters are not loop-corrected by \superlfv, since they may have already been loop-corrected by a spectrum calculator.  That is, \superlfv accepts and uses parameters {\it as is}.

The Higgs fields do not enter in any of the calculations currently offered by \superlfv.  Therefore, their masses and mixing parameters are ignored.

Next, the sfermion mass matrices, along with all other required inputs (\eg, those necessary for left-right mixing and $D$-terms) are supplied by the SLHA spectrum file in the $\overline{\rm DR}$ scheme and used to calculate the running sfermion masses and mixing angles at the {\tt Q=} scale.  Parameters that form the neutralino and chargino mass matrices are read in.  The physical masses and mixings are then calculated.  Wavefunction renormalization of fields is currently not performed in any diagonalization process.

Naively, it would seem simpler to use the sparticle mixing matrices supplied by an SLHA spectrum file.  However, as allowed by the SLHA, those mixings may not be calculated or reported in same manner as the mass matrices.  For example, the masses may be radiatively corrected while the mixing matrices are not.  This matter is very important to calculations involving snuetrinos and charged sleptons, as they tend to include nearly degenerate states.  The orthogonality and near-degeneracy of physical sneutrino and slepton states cause large cancellations within the photon penguin diagrams, hence the parameters used must be sufficiently precise.  Also, insufficient accuracy by an SLHA spectrum calculate may interchange rows in the mixing matrices, leading to garbage results.  Since \superlfv is not tied a specific spectrum calculator, this uncertainty prohibits usage of mixing parameters supplied by SLHA spectrum calculators.  To ensure the integrity of calculations, \superlfv calculates its own physical masses and mixings.  

The SLHA allows for multiple entries of the same {\tt BLOCK} at different scales.  For example, it is common that the {\tt BLOCK GAUGE} is supplied at the GUT scale and soft mass scale.  If \superlfv encounters a block name that has already been read in, the existing block is overwritten; \ie, the block further down the file is retained.  The code checks that the defining scales of Yukawa couplings, gauge couplings, and mass matrices all match.  If they do not match, the code aborts.

Next, using the running masses and mixings evaluated at the soft scale, primitive couplings of mass states (such as snuetrino-lepton-chargino couplings) are evaluated.  Then, also at the soft scale, all relevant effective couplings are then evaluated using only the running masses and primitive couplings. 

Finally, observables are calculated at the soft scale and supplied as output.  In principle, one should run each effective coupling $c$ down to the scale of the initial state lepton mass $m_i$ using a full RG evolution or a leading log approximation, $\Delta c = \beta_{c}\ln(m_i/m_{\rm soft})$.  Since the electromagnetic tensor couplings $A_{2L}^{ij}$ and $A_{2R}^{ij}$ tend to dominate LFV processes in supersymmetric models, \superlfv will RG evolve them from the scale $m_{\rm soft}$ at which they are generated to the muon mass $m_\mu$.  This RG running is performed via the leading-log approximation.  That is,
\beq
A = A \left(1 -\frac{4\alpha}{\pi} \ln\frac{m_{\rm soft}}{m_i} \right)
\eeq
where $A$ represents one of the $A_{2L}^{ij}$ and $A_{2R}^{ij}$ couplings.  In our context, other corrections to other operators are small and neglected.

\subsection{Comparison to existing SLHA tools}

The current version of {\tt SPheno} readily outputs an SLHA spectrum that meets the minimum requirements of \superlfv.\footnote{In principle, other SLHA spectrum calculators may also meet \superlfv's minimal spectrum requirements, however most have omissions in the output -- a trivial obstacle for the authors of SLHA spectrum calculators to resolve.}  Furthermore, {\tt SPheno} currently calculates the branching ratio BR($\mu\to e\gamma$), along with other variants.  This allows for convenient comparison.  It is worth repeating that {\tt SPheno}'s ability to calculate LFV observables is not fully redundant with \superlfv.  A major impetus for \superlfv is to divorce SLHA spectrum calculators from the calculation of LFV observables, allowing for more robust model selection.

Procedural comparisons will be made first.  The electromagnetic dipole couplings $A_{2L}^{ij}$ and $A_{2R}^{ij}$, when squared for observables, significantly lower the observable.  For $\mu\to e\gamma$ and $m_{\rm soft}$ between 100 and 1,000 GeV, the rate BR($\mu\to e\gamma$) is lowered by about 12\% to 17\% \cite{Czarnecki:2001vf}.  {\tt SPheno} currently does not perform this correction.  Also, {\tt SPheno} also performs a loop-corrected wavefunction renormalization of the gaugino and higgsino fields; \superlfv does not.  Neither tool performs a proper matrix treatement for the wavefunction renormalization of the squarks and sleptons.

\begin{table}[t]
\small
\centering
\begin{tabular}{ccccccccc}
&&\superlfv&&&&{\tt SPheno}\\ \cline{2-4} \cline{6-8}
\\[-3mm]
LFV source & BR($\mu\to e\gamma$) & BR($\mu\to 3e$) & ratio && BR($\mu\to e\gamma$) & BR($\mu\to 3e$) & ratio \\[1mm]
\hline\\[-3mm]
$(m^2_{\tilde L})_{12} = (100\,{\rm GeV})^2$ & $4.74\times10^{-12}$ & $3.27\times10^{-14}$ & 0.69\% && $3.57\times10^{-12}$ & $2.19\times10^{-14}$ & 0.61\% \\[1mm]
$(m^2_{\tilde e})_{12} = (100\,{\rm GeV})^2$ & $5.24\times10^{-13}$ & $3.47\times10^{-15}$ & 0.66\% && $1.53\times10^{-13}$ & $8.84\times10^{-16}$ & 0.58\% \\[1mm]
\hline
\end{tabular}
\caption{{\bf Numerical comparison of \superlfv v1.0 and {\tt SPheno} v3.2.1.}  For illustrative purposes, an mSUGRA model augmented by an LFV source is presented.  The mSUGRA model is parameterized as $m_0=2.5$ TeV, $m_{1/2}=1.5$ TeV, $A_0=0$, $\tan\beta=10$, and $\mu>0$.  Here, ``ratio'' is defined as BR($\mu\to 3e$)/BR($\mu\to e\gamma$).}
\label{comparisons}
\end{table}

Numerical comparison examples for \superlfv v1.0 and {\tt SPheno} v3.2.1 are listed in Table \ref{comparisons}.  All discrepancies have been traced back to the neutralino and chargino mixing matrices.  All running couplings, running masses, and sfermion mixing for \superlfv and {\tt SPheno} have been carefully checked to match to several digits of accuracy, typically 1 part in $10^4$ at minimum.  The fact that {\tt SPheno} performs a wavefunction renormalization of the chargino and neutralino matrices while \superlfv does not, creates a small discrepancies in the neutralino and chargino mixing matrices of {\tt SPheno} and \superlfv.  The near-degeneracy of sneutrinos and charged sleptons, along with the orthogonality of their states, results in {\it precise} cancellations in calculating the electromagnetic dipole couplings $A_{2L}^{ij}$ and $A_{2R}^{ij}$.  Small deviations in the neutralino and chargino mixing matrices of {\tt SPheno} and \superlfv will result in imprecise cancellations.  Given that rates are proportional to $|A_{2L}^{ij}|^2 + |A_{2R}^{ij}|^2$, the net effect is that \superlfv's rates are higher than {\tt Spheno}'s but on the same order.  Note that this is only an artifact of using a {\tt SPheno} spectrum file as input for \superlfv.

\begin{figure}[!b]
\begin{center}\includegraphics{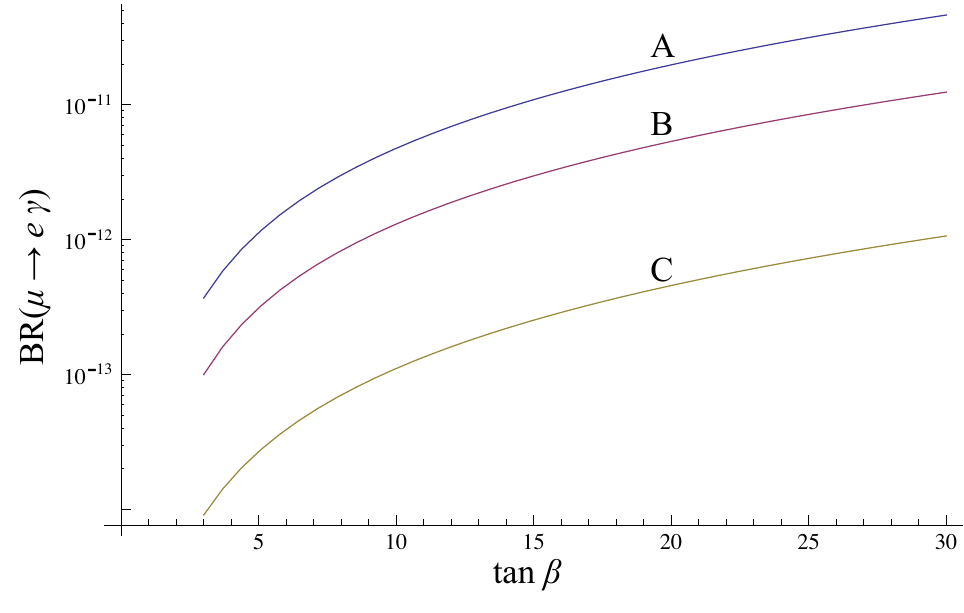}\end{center}
\caption{{\bf A demonstration of the $\tan\beta$ enhancement of $\mu\to e
\gamma$.}  All models (A, B, C) shown have spectra generated by mSUGRA boundary conditions.  In common, they have $A_0=0$ and $\mu>0$.  Model A represents $m_0 = 2$ TeV, $m_{1/2}=1.6$ TeV, and $(m^2_{\tilde L})_{12}=(100\,{\rm GeV})^2$.  Model B represents $m_0 = 500$ GeV, $m_{1/2}=300$ GeV, and $(m^2_{\tilde L})_{12}=(3\,{\rm GeV})^2$.  Model C represents $m_0 = 1$ TeV, $m_{1/2}=800$ GeV, and $(m^2_{\tilde L})_{12}=(10\,{\rm GeV})^2$.}
\label{tanbetaGraph}
\end{figure}

Given a spectrum over which an author has full control, \superlfv's precision is well-controlled.  To demonstrate this, the $\tan\beta$ enhancement of $\mu\to e\gamma$ is demonstrated in Fig.~\ref{tanbetaGraph}.  Recall how this enhancement arises.  The $v_u$ factor of $\tan\beta$ arises from the higgsino-gaugino mass-insertion within the $\mu\to e\gamma$ loop.  The $1/v_d$ factor arises from the muon-higgsino Yukawa coupling.  $v_u$ is encoded into the neutralino (chargino) mixing matrix.  $v_d$ is encoded into the muon-neutralino (chargino) Yukawa couplings, which is composed of elements from the mixing matrices of the slepton (sneutrino) and neutralino (chargino).  Hence, without proper accuracy, the $\tan\beta$ enhancement is easily destroyed.

\section{Installing and operating \superlfv}

\subsection{Installation}

\superlfv is packaged as a single file, named {\tt superlfv-1.0.0.zip}.  It is installed by the following process.
\begin{enumerate}
\item Obtain the \superlfv code from the HepForge distribution service at:
\begin{center}
{\tt http://superlfv.hepforge.org}
\end{center}
\item Double-click {\tt superlfv.zip}. On most modern computing platforms, a {\tt .zip} file will automatically expand into constituent files.  A resulting directory named {\tt SuperLFV} will appear.
\item In a terminal window, change the directory to {\tt SuperLFV}.
\end{enumerate}

\subsection{Immediate usage}
\superlfv is supplied as a single pre-compiled Java JAR file {\tt superlfv.jar} which, aside from an SLHA spectrum file, contains all that is necessary to start using \superlfv immediately.  There is no need to expand the JAR file into constituent files.  As Java code, \superlfv should run on any modern system with a Java Virtual Machine.  An example default input file is supplied as {\tt input.spc}.  To immediately run the code using the default input file, using a command line interface type:

\begin{center}
{\tt java -classpath superlfv.jar superlfv}
\end{center}
This command is tedious and may be vastly simplified, as described below.  By default, only the LFV rates are displayed.  The supplied example  {\tt input.spc} yields the following output.\\

\noindent\hrulefill
{\tt\footnotesize\begin{verbatim}
SuperLFV 1.0

Observable:  BR(l -> l' gamma)
  BR(mu- -> e- gamma) = 1.891993318E-14
  BR(tau- -> e- gamma) = 8.773022513E-36
  BR(tau- -> mu- gamma) = 1.925024531E-34

Observable:  R(mu N -> e N)
  BR(mu Ti -> e Ti) = 1.197657937E-16
  BR(mu Al -> e Al) = 7.041410720E-17

Observable:  BR(l -> l' l' l')
  BR(mu- -> e- e- e+) = 1.311991365E-16
  BR(tau- -> e- e- e+) = 1.734152699E-38
  BR(tau- -> mu- mu- mu+) = 9.690935025E-38
\end{verbatim}}
\noindent\hrulefill

\subsection{Long-term usage}
Invoking \superlfv is vastly simplified by creating an alias.  On Unix variants, one should edit the {\tt {\footnotesize$\sim$}/.profile} file to include the following line.

\begin{center}
{\tt alias superlfv="java -classpath} {\it directory-path}{\tt/superlfv.jar superlfv"}
\end{center}
Here, ``{\it directory-path}'' should be replaced by the directory path that contains the {\tt superlfv.jar} file.  This may be obtained by typing ``{\tt pwd}'' in a terminal window of a Unix variant system.  To invoke the change made to the {\tt {\footnotesize$\sim$}/.profile} file, either open a new terminal window or manually invoke it via ``{\tt source {\footnotesize$\sim$}/.profile}''.  From here on, \superlfv may be invoked with the following simple command:

\begin{center}
{\tt superlfv}
\end{center}
All example commands to invoke \superlfv will use this simplified command.

\begin{table}
\centering
\begin{tabular}{|r|l|}
\hline
{\tt\bf c} & Displays the primitive and effective \underline{\tt\bf c}ouplings used for calculations. \\
{\tt\bf C} & Same as {\tt\bf c} but with the contributions of various amplitude classes shown. \\
{\tt\bf i} & Allows an \underline{\tt\bf i}nput file to be specified. \\
{\tt\bf p} & Displays the \underline{\tt\bf p}arameters (running masses and mixings) used for calculations.\\
{\tt\bf s} & \underline{\tt\bf s}uppresses the output of observables. \\
{\tt\bf w} & Displays the observables \underline{\tt\bf w}ith contributions from various amplitude classes.\\
\hline
\end{tabular}
\caption{\bf A list of command line options.}
\label{options}
\end{table}

\subsection{Options}
Table \ref{options} shows a list options that are supplied in the following format.

\begin{center}
{\tt superlfv} -{\it options} [{\it input file}]
\end{center}
These options may be used in conjunction.  For example, to display the spectrum parameters used for the calculation in addition to supplying an input file different from {\tt input.spc}, one may use:

\begin{center}
{\tt superlfv -pi myinput.spc}
\end{center}
The ordering of options is arbitrary.  Options {\tt -ip} would produce the same results in this example.

One feature of \superlfv is to report quantities broken down into various contributions.  For example, using the option {\tt -w}, the $\mu\to 3e$ branching ratio will explicitly show contributions from photon-exchange, $Z$-exchange, box diagrams, and interference terms.  The same insights may be gained from primitive and effective couplings using the {\tt -C} option.

Output using the {\tt -w} apears in the manner below.\\

\noindent\hrulefill
{\tt\footnotesize\begin{verbatim}
Observable:  BR(l -> l' gamma)
  BR(mu- -> e- gamma) = 1.891993318E-14 = 4.863598412E-16 (neutralino) + 2.547321573E-14 (chargino)
                        + -7.039642383E-15 (interference)
  BR(tau- -> e- gamma) = 8.773022513E-36 = 8.773022513E-36 (neutralino) + 0.00000000000 (chargino)
                        + 0.00000000000 (interference)
  BR(tau- -> mu- gamma) = 1.925024531E-34 = 1.925024531E-34 (neutralino) + 0.00000000000 (chargino)
                        + 0.00000000000 (interference)
\end{verbatim}}
\noindent\hrulefill

\vspace{4mm}
Output using the {\tt -C} apears in the manner below.\\

\noindent\hrulefill
{\tt\footnotesize\begin{verbatim}
Couplings:  lepton-photon-lepton 1-loop vector and tensor couplings
  A_1R^11 = 9.604168151E-09 = 1.283972049E-09 (neutralino) + 8.320196103E-09 (chargino)
  A_1R^12 = -6.564753992E-14 = -1.765258224E-14 (neutralino) + -4.799495768E-14 (chargino)
  A_1R^13 = 8.132763928E-28 = 8.132763928E-28 (neutralino) + 0.00000000000 (chargino)
  A_1R^21 = -6.564753992E-14 = -1.765258224E-14 (neutralino) + -4.799495768E-14 (chargino)
  A_1R^22 = 9.604923201E-09 = 1.284201890E-09 (neutralino) + 8.320721312E-09 (chargino)
  A_1R^23 = -2.353389164E-24 = -2.353389164E-24 (neutralino) + 0.00000000000 (chargino)
\end{verbatim}}
\noindent\hrulefill

\vspace{4mm}
Output using the {\tt -p} apears in the manner below.\\

\noindent\hrulefill
{\tt\footnotesize\begin{verbatim}
SuperLFV 1.0

Scale Q = 1000.0

Higgs parameters:
  tan(beta) = 10.00000000000
  v_u = 242.39690736872
  v_d = 24.23969073687
  v = 243.60587700000
  mu = 399.82391000000

Gauge couplings:
  g_Y = 0.36283193300
  g_2 = 0.64585703200
  g_3 = 1.07837289000

Yukawa couplings:
  y_u = 8.493623820E-06
  y_c = 0.00359564127
  y_t = 0.87077568600
  y_d = 1.364646190E-04
  y_s = 0.00286579735
  y_b = 0.13692728200
  y_e = 2.982727560E-05
  y_mu = 0.00616732139
  y_tau = 0.10376481700

Fermion mixing matrices:
  V_CKM = 
  0.97419203200  0.22569459200  0.00344018712  
  -0.22564536100  0.97335925700  0.04069308400  
  0.00583567101  -0.04041914040  0.99916577100  

  U_PMNS = 
  1.00000000000  7.686990450E-08  0.00000000000  
  -7.686990450E-08  1.00000000000  0.00000000000  
  0.00000000000  0.00000000000  1.00000000000  


Spectrum:  Squarks
  Masses:  
  m2_u1 = 3.662621404E05
  m2_u2 = 5.726255696E05
  m2_u3 = 7.180402438E05
  m2_u4 = 7.180569991E05
  m2_u5 = 7.672395776E05
  m2_u6 = 7.672430938E05
  m2_d1 = 5.248360653E05
  m2_d2 = 6.813001398E05
  m2_d3 = 6.910222744E05
  m2_d4 = 6.910285509E05
  m2_d5 = 7.241148586E05
  m2_d6 = 7.241230542E05

  Mixing matrices:
  U_u = 
  1.971336933E-05  2.332759829E-04  0.38641862210  -5.572092936E-07  -6.437877395E-06  0.92232347560  
  -1.138264522E-04  -0.00134740573  -0.92232253487  2.739826674E-06  2.970168476E-05  0.38641857140  
  -0.07314197920  -0.99728970343  0.00133727625  -1.372091198E-06  -0.00784958826  -3.065230137E-04  
  0.99732153150  -0.07313980192  -1.483096383E-05  1.854684657E-05  -5.758580275E-04  3.391928336E-06  
  -1.848175312E-07  0.00787063396  -4.037200977E-05  6.224076218E-06  -0.99996902521  7.943845986E-06  
  -1.859720289E-05  -5.702575808E-08  2.744681143E-06  0.99999999980  6.223738233E-06  -5.453256277E-07  

  U_d = 
  -0.00564646735  0.03910914089  -0.99675023977  3.251631993E-06  2.938364194E-05  -0.07019651810  
  0.00184956358  -0.01281420614  0.06973474530  -2.145209511E-06  -2.886612624E-04  -0.99748150720  
  -1.397029538E-06  -0.00699365618  -3.289657427E-04  2.846389612E-05  -0.99997542612  3.562264875E-04  
  3.336819281E-04  -6.002136498E-07  1.450037733E-06  0.99999994392  2.846733857E-05  -1.431055699E-06  
  -0.16598733285  0.98522716484  0.04031152804  5.610198104E-05  -0.00690715109  -0.01014434873  
  -0.98610992821  -0.16608670521  -9.472630988E-04  3.289159902E-04  0.00116336550  2.386035123E-04  


Spectrum:  Charged sleptons
  Masses:  
  m2_e1 = 6.052707465E04
  m2_e2 = 6.376245232E04
  m2_e3 = 6.377409414E04
  m2_e4 = 4.160578137E05
  m2_e5 = 4.221118179E05
  m2_e6 = 4.221332878E05

  Mixing matrices:
  U_e = 
  -9.110035956E-19  3.003274668E-15  0.99978452461  4.591381300E-17  1.865954857E-16  0.02075823592  
  0.08621816137  -0.99627553490  4.956493417E-15  5.103829867E-07  -0.00121951244  1.038557432E-16  
  0.99627628127  0.08621809678  -4.274129627E-16  5.897817520E-06  1.055405394E-04  -8.956081394E-18  
  7.633941527E-22  4.531569693E-18  -0.02075823592  -1.251423047E-16  -2.967572004E-16  0.99978452461  
  -3.415966935E-09  -0.00122407083  4.013172943E-17  -3.375193690E-10  0.99999925083  1.456935011E-16  
  5.919859989E-06  1.693127777E-11  2.154393278E-17  -0.99999999998  -3.374799962E-10  -2.541013616E-17  


Spectrum:  Charginos
  Masses:  
  m_C1 = 180.67092936012
  m_C2 = 420.58024290262

  Mixing matrices:
  U = 
  0.93957277159  -0.34234924694  
  0.34234924694  0.93957277159  

  V = 
  0.98331246394  -0.18192470493  
  0.18192470493  0.98331246394  


Spectrum:  Neutralinos
  Masses:  
  m_N1 = 100.87739321550
  m_N2 = 181.06276934770
  m_N3 = -405.56812343355
  m_N4 = 420.17616587035

  Mixing matrices:
  N = 
  -0.98936250240  0.04879451751  -0.12960484767  0.04453669689  
  -0.08376027510  -0.96033420653  0.23511433667  -0.12435303281  
  0.05528440736  -0.08366717812  -0.69738290128  -0.70964817100  
  -0.10530759453  0.26149017215  0.66451555373  -0.69206378265  


Spectrum:  Sneutrinos
  Masses:  
  m2_nu1 = 5.461111497E04
  m2_nu2 = 5.769697773E04
  m2_nu3 = 5.770810200E04

  Mixing matrices:
  U_nu = 
  0.00000000000  0.00000000000  1.00000000000  
  0.09026203273  -0.99591805157  0.00000000000  
  0.99591805157  0.09026203273  0.00000000000  


Trilinears:
  a_u = 
  -0.00499689032  2.262252780E-09  3.249917170E-08  
  9.576890890E-07  -2.11533788000  1.628600610E-04  
  0.00337362315  0.03993516120  -394.28806800000  

  a_d = 
  -0.09843658160  -1.445114440E-06  3.514463130E-05  
  -3.034783470E-05  -2.06698853000  -0.00511193285  
  0.03527477130  -0.24432330100  -92.26523090000  

  a_e = 
  -0.00451416806  0.00000000000  0.00000000000  
  0.00000000000  -0.93336550500  0.00000000000  
  0.00000000000  0.00000000000  -15.61232640000  
\end{verbatim}}
\noindent\hrulefill

\subsection{Minimal parameter inputs}

The minimum requirements for an input file that contains the MSSM couplings and mass parameters are as follows.
\begin{itemize}
\item Gauge and Yukawa couplings must be specified in the SLHA blocks {\tt GAUGE, YU, YD,} and {\tt YE}.
\item Quark and lepton mixings must be specified in the the SLHA blocks {\tt VCKM} and {\tt UPMNS}.
\item Soft mass parameters must be specified in {\tt MSQ2, MSU2, MSD2, MSL2, MSE2, TU, TD, TE} and {\tt MSOFT}.  Only non-zero elements of the mass matrices need be specified.  For {\tt MSOFT}, only the gaugino soft mass parameters $M_1$, $M_2$, and $M_3$ must be entered.
\item Parameters $\mu$, $\tan\beta$, and the Higgs vev $v$ are to be entered in the {\tt HMIX} block.  The Higgs vev may be omitted.  If so, it is approximated by the value obtained by using the current PDG value of the Fermi constant $G_F$.
\end{itemize}
All parameters must be specified at the same scale, using the SLHA {\tt Q=} convention.  An example file {\tt minimal.spc}, provided in the distribution, is shown below.

\noindent\hrulefill
{\tt\footnotesize\begin{verbatim}
# Minimal parameter input template for Super LFV
BLOCK gauge Q=  1.00000000E+03
   1    3.62831933E-01       # g_Y
   2    6.45857032E-01       # g_2
   3    1.07837289E+00       # g_3
BLOCK Yu Q=  1.00000000E+03
  1  1     8.49362382E-06    # y_u
  2  2     3.59564127E-03    # y_c
  3  3     8.70775686E-01    # y_t
BLOCK Yd Q=  1.00000000E+03
  1  1     1.36464619E-04    # y_d
  2  2     2.86579735E-03    # y_s
  3  3     1.36927282E-01    # y_b
BLOCK Ye Q=  1.00000000E+03
  1  1     2.98272756E-05    # y_e
  2  2     6.16732139E-03    # y_mu
  3  3     1.03764817E-01    # y_tau
BLOCK VCKM Q=  1.00000000E+03
   1  1     9.74192032E-01
   1  2     2.25694592E-01
   1  3     3.44018712E-03
   2  1    -2.25645361E-01
   2  2     9.73359257E-01
   2  3     4.06930840E-02
   3  1     5.83567101E-03
   3  2    -4.04191404E-02
   3  3     9.99165771E-01
BLOCK UPMNS Q=  1.00000000E+03
   1  1     1.00000000E+00
   1  2     7.68699045E-08
   2  1    -7.68699045E-08
   2  2     1.00000000E+00
   3  3     1.00000000E+00
BLOCK Tu Q=  1.00000000E+03
   1  1    -4.99689032E-03 
   1  2     2.26225278E-09
   1  3     3.24991717E-08
   2  1     9.57689089E-07
   2  2    -2.11533788E+00
   2  3     1.62860061E-04
   3  1     3.37362315E-03
   3  2     3.99351612E-02
   3  3    -3.94288068E+02
BLOCK Td Q=  1.00000000E+03
   1  1    -9.84365816E-02
   1  2    -1.44511444E-06
   1  3     3.51446313E-05
   2  1    -3.03478347E-05
   2  2    -2.06698853E+00
   2  3    -5.11193285E-03
   3  1     3.52747713E-02
   3  2    -2.44323301E-01
   3  3    -9.22652309E+01
BLOCK Te Q=  1.00000000E+03
   1  1    -4.51416806E-03
   2  2    -9.33365505E-01
   3  3    -1.56123264E+01
BLOCK MSOFT Q=  1.00000000E+03
   1    1.03432998E+02       # M_1
   2    1.93115207E+02       # M_2
   3    5.68025363E+02       # M_3
BLOCK MSL2 Q=  1.00000000E+03
   1  1     6.16982264E+04
   1  2     1.00000000E+00   # The source of LFV lies here.
   2  1     1.00000000E+00
   2  2     6.16872834E+04
   3  3     5.86013300E+04
BLOCK MSE2 Q=  1.00000000E+03
   1  1     4.20218854E+05
   2  2     4.20196836E+05
   3  3     4.13987017E+05
BLOCK MSQ2 Q=  1.00000000E+03
   1  1     7.20764254E+05
   1  2     4.63631320E+01
   1  3    -1.12707907E+03
   2  1     4.63631320E+01
   2  2     7.20449561E+05
   2  3     7.80647919E+03
   3  1    -1.12707907E+03
   3  2     7.80647919E+03
   3  3     5.22563390E+05
BLOCK MSU2 Q=  1.00000000E+03
   1  1     7.68519383E+05
   1  2     1.65077717E-09
   1  3     5.78795005E-06
   2  1     1.65077717E-09
   2  2     7.68512439E+05
   2  3     2.90046080E-02
   3  1     5.78795005E-06
   3  2     2.90046080E-02
   3  3     3.76076532E+05
BLOCK MSD2 Q=  1.00000000E+03
   1  1     6.90390410E+05
   1  2    -7.61272825E-07
   1  3     8.84248145E-04
   2  1    -7.61272825E-07
   2  2     6.90385750E+05
   2  3    -1.28617580E-01
   3  1     8.84248145E-04
   3  2    -1.28617580E-01
   3  3     6.79889912E+05
BLOCK HMIX Q=  1.00000000E+03
   1    3.99823910E+02       # mu
   2    1.00000000E+01       # tan(beta)(Q)
   3    2.43605877E+02       # v(Q)
\end{verbatim}
} 
\noindent\hrulefill

\section{Summary and outlook}

Presently we are in an era in which MEG, a dedicated $\mu\to e\gamma$ search, is currently acquiring data, along with an operational LHC, an unprecedented variety of future non-collider experiments, and future B-factories.  Accordingly, a need for rapid model-discrimination arises.  We have introduced \superlfv, a new SLHA tool for calculating low-energy lepton flavor violating observables in the context of supersymmetric models.  \superlfv offers a few unique features, including (1) independence from existing SLHA spectrum calculators, (2) the ability to accept an SLHA spectrum file generated by an existing specturm calculator, personal code, or by hand; and (3) an ability to report observables and couplings as a sum of contributions.  Features (1) and (2) allow arbitrary model selection.  Feature (3) promotes analytical insights.

\superlfv is intended to be a well-supported addition to the SLHA library.  The initial offering is minimally comprehensive.  Future enhancements will be demand-driven and may include greater accuracy (\eg, wavefunction renormalization of the neutralino and chargino fields, more sophistication in handling the nuclear aspects of coherent muon conversion, {\it etc.}), the inclusion of neglected amplitudes (\eg, Higgs-mediated LFV), more observables, or possibly expansion beyond the MSSM.

\section{Acknowledgements}
The author would like to express his gratitude to J.~Wells and K.~Tobe for their insight and support throughout this project.  This project was developed over three years and work was performed on it at many institutions.  The author thanks the members of UC Irvine, UC Davis, the University of Hawaii, the University of Michigan, Northeastern University, Nagoya University, the CERN Theory Group, the Argonne HEP Theory Group, Indiana University, and the Brookhaven High Energy Theory Group for providing productive environments and engaging conversations.  This work was supported by National Science Awards awards PHY-1068420 and PHY-1068420, the Rhode Island College (RIC) Faculty Research Fund, and the RIC Faculty Development Fund.

The \superlfv uses public code from the Apache Commons library for some mathematical operations.

\appendix
\section{MSSM couplings}

In the following, sfermion indices are denoted as $I$, fermion indices are $j$, and neutralino (chargino) indices are $A$.  In 4-component Dirac notation, the interactions of a sfermion-fermion-neutralino are summarized below.
\bea
-\lag_{\tilde u u \tilde\chi^0} &\equiv& \tilde u_I \bar u_j (n_{IjA}^{u_L} P_L + n_{IjA}^{u_R} P_R) \tilde\chi_A^0 + {\rm h.c.} \\
-\lag_{\tilde d d \tilde\chi^0} &\equiv& \tilde d_I \bar d_j (n_{IjA}^{d_L} P_L + n_{IjA}^{d_R} P_R) \tilde\chi_A^0 + {\rm h.c.} \\
-\lag_{\tilde e e \tilde\chi^0} &\equiv& \tilde e_I \bar e_j (n_{IjA}^{e_L} P_L + n_{IjA}^{e_R} P_R) \tilde\chi_A^0 + {\rm h.c.} \\
-\lag_{\tilde \nu \nu \tilde\chi^0} &\equiv& \tilde\nu_i \bar\nu_{Lj} (n_{ijA}^{\nu_L} P_L + n_{ijA}^{\nu_R} P_R) \tilde\chi_A^0 + {\rm h.c.}
\eea
\bea
n_{IjA}^{u_L} &\equiv& \sqrt 2 g_Y Y_{\bar u} N_{A1}^* U_{\tilde u}^{I(j+3)*} + y_u^{ij} N_{A4}^* U_{\tilde u}^{Ii*} \\
n_{IjA}^{u_R} &\equiv& \sqrt 2 g_Y Y_Q N_{A1}^* U_{\tilde u}^{Ij*} + \sqrt 2 g_2 t^3_{u_L} N_{A2}^* U_{\tilde u}^{Ij*} + y_u^{ij*} N_{A4} U_{\tilde u}^{I(i+3)*} \\[3mm]
n_{IjA}^{d_L} &\equiv& \sqrt 2 g_Y Y_{\bar d} N_{A1}^* U_{\tilde d}^{I(j+3)*} + y_d^{ij} N_{A3}^* U_{\tilde d}^{Ii*} \\
n_{IjA}^{d_R} &\equiv& \sqrt 2 g_Y Y_Q N_{A1}^* U_{\tilde d}^{Ij*} + \sqrt 2 g_2 t^3_{d_L} N_{A2}^* U_{\tilde d}^{Ij*} + y_d^{ij*} N_{A3} U_{\tilde d}^{I(i+3)*} \\[3mm]
n_{IjA}^{e_L} &\equiv& \sqrt 2 g_Y Y_{\bar e} N_{A1}^* U_{\tilde e}^{I(j+3)*} + y_e^{ij} N_{A3}^* U_{\tilde e}^{Ii*} \\
n_{IjA}^{e_R} &\equiv& \sqrt 2 g_Y Y_L N_{A1}^* U_{\tilde e}^{Ij*} + \sqrt 2 g_2 t^3_{e_L} N_{A2}^* U_{\tilde e}^{Ij*} + y_e^{ij*} N_{A3} U_{\tilde e}^{I(i+3)*} \\[3mm]
n_{ijA}^{\nu_L} &\equiv& 0 \\
n_{ijA}^{\nu_R} &\equiv& \sqrt 2 g_Y Y_L N_{A1}^* U_{\tilde\nu}^{ij*} + \sqrt 2 g_2 t^3_{\nu_L} N_{A2}^* U_{\tilde\nu}^{ij*}
\eea

The sfermion-fermion-chargino interactions are as follows.
\bea
-\lag_{\tilde d u \tilde\chi} &\equiv& \tilde d_I \bar u_j(c^{u_L}_{IjA} P_L + c^{u_R}_{IjA} P_R) \tilde\chi^+_A + {\rm h.c.} \\
-\lag_{\tilde u d \tilde\chi} &\equiv& \tilde u_I \bar d_j(c^{d_L}_{IjA} P_L + c^{d_R}_{IjA} P_R) \tilde\chi^-_A + {\rm h.c.} \\
-\lag_{\tilde \nu e \tilde\chi} &\equiv& \tilde \nu_i \bar e_j (c^{e_L}_{ijA} P_L + c^{e_R}_{ijA} P_R) \tilde\chi^-_A + {\rm h.c.} \\
-\lag_{\tilde e \nu \tilde\chi} &\equiv& \tilde e_I \bar\nu_{Lj} (c^{\nu_L}_{IjA} P_L + c^{\nu_R}_{IjA} P_R) \tilde\chi^+_A + {\rm h.c.} 
\eea
\bea
c_{IjA}^{u_L} &\equiv& -y_u^{ij} V_{A2}^* U_{\tilde d}^{Ii*} \\
c_{IjA}^{u_R} &\equiv& g_2 U_{A1} U_{\tilde d}^{Ij*} - y_d^{ij*} U_{A2} U_{\tilde d}^{I(i+3)*} \\[3mm]
c_{IjA}^{d_L} &\equiv& -y_d^{ij*} U_{A2}^* U_{\tilde u}^{Ii*} \\
c_{IjA}^{d_R} &\equiv& g_2 V_{A1} U_{\tilde u}^{Ij*} - y_u^{ij} U_{A2}^* U_{\tilde u}^{I(i+3)*} \\[3mm]
c_{ijA}^{e_L} &\equiv&  -y_e^{kj*} U_{A2}^* U_{\tilde\nu}^{ik*} \\
c_{ijA}^{e_R} &\equiv&  g_2 V_{A1} U_{\tilde \nu}^{ij*}
\\[3mm]
c_{IjA}^{\nu_L} &\equiv& 0 \\
c_{IjA}^{\nu_R} &\equiv& g_2 U_{A1} U_{\tilde e}^{Ij*} - y_e^{ij*} U_{A2} U_{\tilde e}^{I(i+3)*}
\eea

From here, one may calculate the effective couplings.  The electromagnetic form factors of eqn.~(\ref{emOperators}) are generated at one-loop order by neutralinos and charginos.  For muon decay, these operators are to be evaluated at the muon mass scale, \ie~$A_{2R}^{ij}(q^2 = m_\mu^2)$.  Following \cite{Hisano:1995cp}, electromagnetic form factors couplings are the following.

\bea
A_{1L}^{ij} &=& \frac 1 {576\pi^2} \frac {n^{e_R*}_{IiA} n^{e_R}_{IjA}} {m^2_{\tilde e_I}} \frac {2 - 9x + 18x^2 - 11x^3 + 6x^3\ln x}{(1-x)^4} \nonumber\\
&&- \frac 1{576\pi^2} \frac {c^{e_R*}_{IiA} c^{e_R}_{IjA}} {m^2_{\tilde\nu_I}} \frac {16 - 45y + 36y^2 -7y^3 + 6(2-3y)\ln y}{(1-y)^4} \\
A_{1R}^{ij} &=& A_{1L}^{ij}|_{L\leftrightarrow R} \\
A_{2L}^{ij} &=& \frac 1 {32\pi^2} \frac {n^{e_L*}_{IiA} n^{e_L}_{IjA}} {m^2_{\tilde e_I}} \frac {1 -6x + 3x^2 + 2x^3 - 6x^2\ln x}{6(1-x)^4} \nonumber\\
&&+ \frac 1 {32\pi^2} \frac {n^{e_R*}_{IiA} n^{e_L}_{IjA} m_{\tilde\chi^0_A}} {m_{e_i} m^2_{\tilde e_I}} \frac {1 - x^2 + 2x \ln x}{(1-x)^3} \nonumber\\
&&- \frac 1 {32\pi^2} \frac {c^{e_L*}_{IiA} c^{e_L}_{IjA}} {m^2_{\tilde\nu_I}} \frac {2 + 3y - 6y^2 + y^3 + 6y\ln y} {6(1-y)^4} \nonumber\\
&& - \frac 1 {32\pi^2} \frac {c^{e_R*}_{IiA} c^{e_L}_{IjA} m_{\tilde\chi^-_A}} {m_{e_i} m^2_{\tilde\nu_I}} \frac {(-3 + 4y - y^2 - 2\ln y)}{(1-y)^3} \\
A_{2R}^{ij} &=& A_{2L}^{ij}|_{L\leftrightarrow R}
\eea
Above, $x \equiv x_{IA} = m_{\tilde\chi^{0}_A}^2 / m^2_{\tilde e_I}$ and $y \equiv y_{IA} = m_{\tilde\chi^{0}_A}^2 / m^2_{\tilde e_I}$.

The effective couplings for $Z$-exchange are
\bea
F_L^{ij} &=& \frac 1 {32\pi^2} n^{e_R*}_{IiA} n^{e_R}_{IjB} (N_{A3} N_{B3} - N_{A4} N_{B4}) (F_{IAB} + 2G_{IAB}) \nonumber\\
&&- \frac 1 {32\pi^2} c^{e_R*}_{IiA} c^{e_R}_{IjB} \left(\frac 12 V_{A2} V_{B2} F_{IAB} - U_{A2} U_{B2} G_{IAB} \right)\\
F_R^{ij} &=& \frac 1 {32\pi^2} n^{e_L*}_{IiA} n^{e_L}_{IjB} (N_{A3} N_{B3} - N_{A4} N_{B4}) (F_{IAB} + 2G_{IAB})\eea
where
\bea
F_{IAB} &=& \ln x_{IA} + \frac 1 {x_{IA} - x_{IB}}
\left(
\frac {x_{IA}^2 \ln x_{IA}} {1 - x_{IA}} - \frac{x_{IB}^2 \ln x_{IB}} {1 - x_{IB}}
\right) \\
{\rm and} \,\,\,
G_{IAB} &=& \frac {m_{\tilde\chi_A} m_{\tilde\chi_B}} {m^2_{\tilde l_I}}
\frac 1 {x_{IA} - x_{IB}}
\left(
\frac {x_{IA} \ln x_{IA}} {1 - x_{IA}} - \frac{x_{IB} \ln x_{IB}} {1 - x_{IB}}
\right) .
\eea
Here, $m^2_{\tilde l_I}$ is $m^2_{\tilde e_I}$ when used in the neutralino contribution (marked by $n_{IiA}$ couplings) or  $m^2_{\tilde\nu_I}$ when used in the chargino contribution (marked by $c_{IiA}$ couplings).

For the box diagrams of $e_i \to 3e_j$, the effective couplings are as follows \cite{Hisano:1995cp}.
\bea
e^2 B_{1L}^{ij} &=& \frac 12 n^{e_R*}_{IiA} n^{e_R}_{JjA} n^{e_R*}_{JjB} n^{e_R}_{IjB} J_{4(ABIJ)}
+ n^{e_R*}_{IiA} n^{e_R*}_{JjA} n^{e_R}_{JjB} n^{e_R}_{IjB} \cdot m_{\tilde\chi^0_A} m_{\tilde\chi^0_B} I_{4(ABIJ)} \nonumber\\
&& + \frac 12 c^{e_R*}_{IiA} c^{e_R}_{JjA} c^{e_R*}_{JjB} c^{e_R}_{IjB} J_{4(ABIJ)} \\
e^2 B_{2L}^{ij} &=& \frac 14 (n^{e_R*}_{IiA} n^{e_R}_{JjA} n^{e_L*}_{JjB} n^{e_L}_{IjB}
+ n^{e_R*}_{IiA} n^{e_L*}_{JjA} n^{e_R}_{JjB} n^{e_L}_{IjB}
- n^{e_R}_{IiA} n^{e_L*}_{JjA} n^{e_L*}_{JjB} n^{e_R}_{IjB}) J_{4(ABIJ)} \nonumber\\
&&- \frac 12 n^{e_R*}_{IiA} n^{e_L}_{JjA} n^{e_L*}_{JjB} n^{e_R}_{IjB} \cdot m_{\tilde\chi^0_A} m_{\tilde\chi^0_B} I_{4(ABIJ)} \nonumber\\
&&+ \frac 14 c^{e_R*}_{IiA} c^{e_R}_{JjA} c^{e_L*}_{JjB} c^{e_L}_{IjB} J_{4(ABIJ)}
- \frac 12 c^{e_R*}_{IiA} c^{e_L}_{JjA} c^{e_L*}_{JjB} c^{e_R}_{IjB} \cdot m_{\tilde\chi^-_A} m_{\tilde\chi^-_B}I_{4(ABIJ)} \\
e^2 B_{3L}^{ij} &=& (n^{e_R*}_{IiA} n^{e_L}_{JjA} n^{e_R*}_{JjB} n^{e_L}_{IjB}
+ \frac 12 n^{e_R*}_{IiA} n^{e_R*}_{JjA} n^{e_L}_{JjB} n^{e_L}_{IjB}) \cdot m_{\tilde\chi^0_A} m_{\tilde\chi^0_B} I_{4(ABIJ)} \nonumber\\
&&+ c^{e_R*}_{IiA} c^{e_L}_{JjA} c^{e_R*}_{JjB} c^{e_L}_{IjB} \cdot m_{\tilde\chi^-_A} m_{\tilde\chi^-_B} I_{4(ABIJ)} \\
e^2 B_{4L}^{ij} &=& \frac 18 n^{e_R*}_{IiA} n^{e_R*}_{JjA} n^{e_L}_{JjB} n^{e_L}_{IjB} \cdot m_{\tilde\chi^0_A} m_{\tilde\chi^0_B} I_{4(ABIJ)} \\
B_{aR}^{ij} &=& B_{aL}^{ij}|_{L \leftrightarrow R}\,\,\,\,{\rm where}\,\,a\in\{1,2,3,4\}
\eea
The loop integrals $I_4$ and $J_4$ are
\bea
iI_{4(ABIJ)} &=& \int \frac{d^4k}{(2\pi)^4} \frac 1
{(k^2 - m_{\tilde\chi_A}^2) (k^2 - m_{\tilde\chi_B}^2) (k^2 - m^2_{\tilde l_I}) (k^2 - m^2_{\tilde l_J}) } ,\\
iJ_{4(ABIJ)} &=& \int \frac{d^4k}{(2\pi)^4} \frac {k^2}
{(k^2 - m_{\tilde\chi_A}^2) (k^2 - m_{\tilde\chi_B}^2) (k^2 - m^2_{\tilde l_I}) (k^2 - m^2_{\tilde l_J}) }.
\eea

The box diagrams of muon conversion have the following effective couplings \cite{Hisano:1995cp}.
\bea
D^{q_L} &=& D^{q_L}_n + D^{q_L}_c \\
e^2 D^{q_L}_n &=& \frac 18 (n^{e_R*}_{I2A} n^{e_R}_{I1B} n^{q_R}_{JqA} n^{q_R*}_{JqB} - n^{e_R*}_{I2A} n^{e_R}_{I1B} n^{q_L*}_{JqA} n^{q_L}_{JqB}) J_{4(ABIJ)} \nonumber\\
&&- \frac 14 (n^{e_R*}_{I2A} n^{e_R}_{I1B} n^{q_L}_{JqA} n^{q_L*}_{JqB}
- n^{e_R*}_{I2A} n^{e_R}_{I1B} n^{q_R*}_{JqA} n^{q_R}_{JqB}) \cdot m_{\tilde\chi^0_A} m_{\tilde\chi^0_B} I_{4(ABIJ)} \\
D_{n}^{q_R} &=& D_{n}^{q_L}|_{L \leftrightarrow R}\,\,\,\,{\rm where}\,\,a\in\{1,2,3,4\} \\
e^2 D^{u_L}_c &=& -\frac 18 c^{e_R*}_{I2A} c^{e_R}_{I1B} c^{u_L}_{J1A} c^{u_L*}_{J1B} J_{4(ABIJ)}
+ \frac 14 c^{e_R*}_{I2A} c^{e_R}_{I1B} c^{u_R*}_{J1A} c^{u_R}_{J1B} \cdot m_{\tilde\chi^-_A} m_{\tilde\chi^-_B} I_{4(ABIJ)} \\
e^2 D^{d_L}_c &=& \frac 18 c^{e_R*}_{I2A} c^{e_R}_{I1B} c^{d_R}_{J1A} c^{d_R*}_{J1B} J_{4(ABIJ)}
- \frac 14 c^{e_R*}_{I2A} c^{e_R}_{I1B} c^{d_L}_{J1A} c^{d_L*}_{J1B} \cdot m_{\tilde\chi^-_A} m_{\tilde\chi^-_B} I_{4(ABIJ)}
\eea

\end{document}